\documentclass[hyper]{prop2015}
\usepackage[english]{babel}

\usepackage{amsfonts,amsthm,amstext,array}
\usepackage{mathrsfs}
\usepackage[arrow,matrix,curve]{xy}

\usepackage{hhline}

\usepackage{hyperref}
\usepackage{graphics}
\usepackage{amsmath}
\usepackage{amssymb,amscd}
\usepackage[arrow,matrix,curve]{xy}
\usepackage{graphicx}
\usepackage{bbm}

\usepackage{pdfsync}

\usepackage{epsfig}

\theoremstyle{plain}
\newtheorem{theo}{Theorem}[section]

\def\={\ =\ }
\def\dd{{\rm d}}

\def\e{{\,\rm e}\,}

\def\ii{{\,{\rm i}\,}}

\newcommand{\bbr}{\mathbbm{R}}

\newcommand{\bbc}{\mathbb{C}}

\newcommand{\call}{\mathcal{L}}

\newcommand{\cals}{\mathcal{S}}
\newcommand{\cali}{\mathcal{I}}
\newcommand{\calo}{\mathcal{O}}
\newcommand{\calm}{\mathcal{M}}

\newcommand{\calp}{\mathcal{P}}

\def\hil{{\mathcal H}}

\def\cG{{\mathcal G}}

\newcommand{\bbz}{{\mathbb Z}}

  \definecolor{violet}{rgb}{.7,0,1}

  \definecolor{dgreen}{rgb}{.2,.5,.1}

  \definecolor{test}{rgb}{.5,1,0}

\newcommand{\remark}[1]{}     				
\category{Proceedings}
\subtitle{\href{http://www.maths.dur.ac.uk/lms/109/index.html}{LMS/EPSRC Durham Symposium on Higher Structures in M-Theory}}
\title{Quantization of Magnetic Poisson Structures}
\author[R.\,J. Szabo]{Richard J. Szabo\inst{a,}\footnote{Corresponding author e-mail:~\href{mailto:R.J.Szabo@hw.ac.uk}{\textsf{R.J.Szabo@hw.ac.uk}}}}
\address[1]{Department of Mathematics, Heriot--Watt University, Edinburgh EH14 4AS, United Kingdom;
{Maxwell Institute for Mathematical Sciences, Edinburgh, United Kingdom};
{The Higgs Centre for Theoretical Physics, Edinburgh, United Kingdom}}

\begin{acknowledgements}
This article is based in part on work done in collaboration with
Severin Bunk, Vladislav Kupriyanov, Dionysios Mylonas, Lukas M\"uller
and Peter Schupp, who we warmly thank for many productive
conversations. We thank Damien Calaque, Alberto Cattaneo and Jouko
Mickelsson for helpful discussions and correspondence, and Branislav
Jur\v{c}o, Christian S\"amann, Urs Schreiber and Martin Wolf for the
invitation to speak at the LMS/EPSRC/Durham Symposium on `Higher Structures
in M-Theory'.
This work was supported by the COST Action MP1405 QSPACE, funded by the
European Cooperation in Science and Technology (COST), and by the Consolidated Grant ST/P000363/1 from the UK Science and Technology
Facilities Council (STFC).
\end{acknowledgements}

\begin{abstract}
We describe three perspectives on higher quantization, using the
example of magnetic Poisson structures which embody recent discussions
of nonassociativity in quantum mechanics with magnetic monopoles and string
theory with non-geometric fluxes. We survey approaches based on
deformation quantization of twisted Poisson structures,
symplectic realization of almost symplectic structures, and geometric quantization using 2-Hilbert
spaces of sections of suitable bundle gerbes. We compare and contrast
these perspectives, describing their advantages and shortcomings in
each case, and mention many open avenues for investigation.
\end{abstract}
\shortabstract
\begin{document}
\maketitle

\section{Introduction}

In this contribution we will study three perspectives on the problem
of higher quantization, which at present is a problem not understood
to the level of ordinary quantization schemes such as geometric
quantization or, more concretely, canonical quantization in quantum
mechanics. We do not attempt a general discussion of the problem; see
e.g.~\cite{Saemann:2012ab,Bunk:2016rta,Bunk:2016gus} for detailed overviews of the issues surrounding higher
quantization generically. Instead, we focus on a particularly simple and tractable
class of models of physical significance, that we call `magnetic
Poisson structures', in which precise statements and advancements can
be made from both mathematical and physical perspectives. These are special examples of non-degenerate twisted Poisson, or
equivalently almost symplectic, structures which are related to recent
discussions of nonassociativity in quantum
mechanics and in non-geometric string theory. 

We define and motivate
the relevant structures in Section~\ref{sec:MPS}, and describe the general
quantization problem involved. We then offer three perspectives on how
to tackle this quantization problem, starting with the most concrete
framework and ending with the most abstract one. Section~\ref{sec:PI}
reviews the well-known approach through deformation quantization. In
Section~\ref{sec:PII} we describe a new approach through an extension
of the well-known procedure of symplectic realization in Poisson
geometry to the case of twisted (or more generally quasi-) Poisson
structures. Finally, in Section~\ref{sec:PIII} we describe an approach
based on a higher version of geometric quantization which brings the
formalism of higher structures into full play by regarding `higher
quantum states' as sections of a suitable bundle gerbe, the
appropriate higher analog of the line bundles usually employed in
ordinary geometric quantization. Each of these approaches has their
own advantages, but also several deficiencies which we explain in detail in
the following.

\section{Magnetic Poisson structures\label{sec:MPS}}

We begin by defining and motivating the specific twisted Poisson
structures that we will attempt to quantize. We shall then describe the specifics
of the quantization problem we wish to address.

\subsection{Definition\label{subsec:MPSdef}}

We work in the simple setting of the $d$-dimensional vector space
$M=\bbr^d$, which we will refer to as `configuration space' in the
following; local coordinates on $M$ are denoted $x$.  We write $M^*$
for the dual vector space and call it `momentum space', with local
coordinates $p$. The (trivial) cotangent bundle $\calm=T^*M=M\times
M^*$ is called `phase space'; it has local coordinates $X=(x,p)$
and the canonical symplectic form 
\begin{align}
\sigma_0(X,X')=p\cdot x'-p'\cdot x \ ,
\end{align}
where a dot denotes the canonical duality pairing between vectors and
covectors.

We fix a (not necessarily closed) two-form $\rho\in\Omega^2(M)$ on configuration space
$M\subset \calm$ and call it a `magnetic field',
for reasons that will become clear from the applications we describe
below. It deforms the symplectic structure $\sigma_0$ to an almost symplectic form
\begin{align}
\sigma_\rho=\sigma_0-\rho \ ,
\label{eq:sigmarho}\end{align}
which is always non-degenerate (because $\sigma_0$ is) but is closed
if and only if $\rho$ is closed; in \eqref{eq:sigmarho} we of course
mean the pullback of $\rho$ under the cotangent bundle projection
$\calm\longrightarrow M$, and we shall frequently abuse notation in this way since
all considerations in the following occur in this simple topologically
trivial setting. Its inverse
$\theta_\rho=\sigma_\rho^{-1}$ gives a bivector which defines the
\emph{magnetic Poisson al\-gebra}  on $C^\infty(\calm)$ with brackets
\begin{align}
\{f,g\}_\rho = \theta_\rho(\dd f\wedge \dd g)
\label{eq:fgrho}\end{align}
for smooth complex-valued phase space functions $f,g\in C^\infty(\calm)$. In particular, for
the coordinate functions $x^i(X)=x^i$ and $p_i(X)=p_i$, with
$i=1,\dots,d$, one has 
\begin{align}
\{x^i,x^j\}_\rho&=0 \ , \nonumber \\[4pt]
\{x^i,p_j\}_\rho&=\delta^i{}_j \ , \label{eq:coordrho} \\[4pt]
\{p_i,p_j\}_\rho&= -\rho_{ij}(x) \ , \nonumber 
\end{align}
where $\rho=\frac12\, \rho_{ij}(x)\,\dd x^i\wedge\dd x^j$ (with
implicit summation over repeated upper and lower indices always understood).

These brackets do not generally define a Poisson algebra but rather an 
\emph{$H$-twisted Poisson structure} on $\calm$, with twisting given
by the three-form $H=\dd\rho$ on $M$ that we shall call a `magnetic
charge'. This means that the Schouten bracket of the bivector
$\theta_\rho$ with itself, which governs the associativity of the
brackets defined by \eqref{eq:fgrho}, is given by the trivector
\begin{align}
[\theta_\rho,\theta_\rho]_{\rm
  S}=\mbox{$\bigwedge^3$}\theta_\rho^\sharp(\dd\sigma_\rho) \ ,
\end{align}
where
$\theta_\rho^\sharp$ denotes the natural contraction of forms to vectors by
the non-degenerate bivector $\theta_\rho$. It vanishes if
and only if $H=0$, while it generically gives a nonassociative algebra
with Jacobiators 
\begin{align}
\{f,g,h\}_\rho :=&\,
                 \frac13\,\big(\{f,\{g,h\}_\rho\}_\rho -\{\{f,g\}_\rho,h\}_\rho
                 -\{g,\{f,h\}_\rho\}_\rho\big) \nonumber \\[4pt]
=&\, [\theta_\rho,\theta_\rho]_{\rm S}(\dd f\wedge \dd
g\wedge \dd h) \ .
\label{eq:fghrho}\end{align}
In particular, on coordinate functions the only non-vanishing
Jacobiators are given by 
\begin{align}
\{p_i,p_j,p_k\}_\rho=-H_{ijk}(x) \ ,
\end{align}
where $H=\frac1{3!}\,H_{ijk}(x)\, \dd x^i\wedge\dd x^j\wedge\dd x^k$.
This is a generalization of the nonassociative algebras that were
introduced in the physics literature of the 1980's in the context of
the G\"unaydin--Zumino model~\cite{Gunaydin:1985ur}, and they are a
natural playground for higher structures as we explain in the following. 

There are obvious generalizations of this model that one can
envisage. One can of course replace $M$ with any manifold (not
necessarily a vector space). The cotangent bundle $\calm=T^*M$ can likewise
be replaced by any Lie algebroid $\call$ over $M$, with
$\rho\in\Gamma\big(M,\bigwedge^2\call^*\big)$ a Lie algebroid two-cochain
which defines a central extension of $\call$.
Then $H=\dd\rho$ defines a Lie algebroid three-cocycle with values in
the kernel of the anchor map of the extension and hence can be used
to endow $\call$ with the structure of an $H$-twisted Lie
algebroid~\cite{Grutzmann:1005.5680}, which is in particular a 2-term $L_\infty$-algebroid. However, here we stick to
this concrete and simple example as it will capture the essential
features that we wish to describe in this contribution. Let us first
briefly 
explain two motivations from physics for being interested in the
quantization of such structures.

\subsection{Application I: magnetic monopoles\label{sec:monopole}}

Our first application is to quantum mechanics. The magnetic Poisson structure in $d=3$ dimensions with
$\rho_{ij}=e\,\varepsilon_{ijk}\, B^k$ governs the motion of an
electric charge $e$ in a magnetic field $\vec B$ on $\bbr^3$. When the
two-form $\rho$ is closed, $\dd\rho=0$, this corresponds to the
classical Maxwell theory without magnetic sources,
${\rm div}\ \vec B=0$, and with a globally defined magnetic vector
potential $\vec A$ on $\bbr^3$ such that $\vec B={\rm curl}\ \vec
A$. This is the case in which the magnetic Poisson
structure yields an associative Poisson algebra. 

The simplest instance of nonassociativity comes from Dirac's
modification of Maxwell's theory, which considers a singular
delta-function distributed source of magnetic charge at the origin
$\vec0$ of $\bbr^3$ representing a Dirac monopole. The magnetic field
sourced by the Dirac monopole on $\bbr^3\setminus\{\vec0\}$ is given by
\begin{align}
\vec B_{\rm D}=g\,\frac{\vec x}{|\vec
  x|^3}={\rm curl}\ \vec A_{\rm D} \ ,
\label{eq:Diracmonopole}\end{align}
where $g$ is the magnetic charge and the local magnetic vector
potential
\begin{align}
\vec A_{\rm D}\=\frac{g}{|\vec x|} \, \frac{\vec x\times\vec n}{|\vec x|-\vec x\cdot \vec n}
\label{eq:Diracpotential}\end{align}
has an additional Dirac string singularity along the semi-infinite line emanating
from the location of the monopole at $\vec0$ in the
direction of a fixed unit vector $\vec n$. This makes a modest
connection between higher structures and experiment, in that analog
systems of Dirac monopoles have been observed through neutron
scattering events off of spin ice pyrochlore lattices, see
e.g.~\cite{Castelnovo:2007qi,Morris:2010ma} for early reports. These
lattices have tetrahedral atomic arrangements with magnetic dipoles
through the corners of the tetrahedra, and local magnetic pole defects in
the lattice can be
observed in interference patterns from interactions with
neutrons, which themselves have a dipole moment, and an external
magnetic field. In this
sense Dirac strings and monopoles arise as emergent states of matter; see~\cite{Szabo:2017yxd} for a more detailed
discussion in the present context and further references.

In this contribution we are ultimately interested in the cases where
the twisting three-form $H=\dd\rho\neq0$ is non-singular and describes smooth
distributions of magnetic charge. Although not yet visible in
experiment, one can regard such distributions as arising in an effective framework where we treat a
system of Dirac monopoles in a long wavelength limit at scales much
larger than the lattice spacing used in realistic scenarios. In such a
setting one must account for nonassociativity along smooth
submanifolds, and not just at the point supports of Dirac monopoles. Then a foundational question about the quantum
dynamics of electric charge in such distributions arises: 
\begin{itemize}
\item What is a sensible
framework for \emph{nonassociative quantum mechanics}?
\end{itemize}
By `sensible' we mean a formalism that agrees with the usual physical
requirements of a quantum theory and which has the potential to be
experimentally tested.

\subsection{Application II: locally non-geometric fluxes\label{sec:Rflux}}

A somewhat more speculative application is to string theory. Consider
the \emph{magnetic duality} transformation $(x,p)\longmapsto (p,-x)$
on the phase space $\calm$ (for any $d$), which preserves the canonical
symplectic structure $\sigma_0$. In this case we can trade the
configuration space two-form $\rho\in\Omega^2(M)$ with a momentum
space two-form $\beta\in\Omega^2(M^*)$ such that the twisted Poisson
brackets among coordinate functions become
\begin{align}
\{x^i,x^j\}_\beta&=-\beta^{ij}(p) \ ,\notag \\[4pt]
  \{x^i,p_j\}_\beta&=\delta^i{}_j \ ,  \\[4pt] 
  \{p_i,p_j\}_\beta&=0 \ ,\notag
\end{align}
where $\beta=\frac12\,\beta^{ij}(p)\, \dd p_i\wedge\dd p_j$.
In this case the twisting three-form is the `$R$-flux'
$R=\dd\beta\in\Omega^3(M^*)$ which describes a nonassociative
configuration space with the non-vanishing Jacobiator among coordinate
functions given by
\begin{align}
\{x^i,x^j,x^k\}_\beta = -R^{ijk}(p) \ ,
\end{align}
where $R=\frac1{3!}\, R^{ijk}(p)\,\dd p_i\wedge\dd p_j\wedge\dd
p_k$. This dynamical system is called the \emph{$R$-flux model}, and
it conjecturally describes the phase space of closed strings
propagating in `locally non-geometric' $R$-flux backgrounds, see
e.g.~\cite{Blumenhagen:2010hj,Lust:2010iy,Blumenhagen:2011ph,Mylonas:2012pg,Freidel:2017nhg}. In
this example another foundational question arises: 
\smallskip
\begin{enumerate}[i)]
\item What substitutes for canonical quantization of locally non-geometric closed strings?
\end{enumerate}

\subsection{Magnetic translation operators\label{sec:Magtransl}}

Let us now describe the quantization problem for magnetic
Poisson structures in some generality. Quantization should be a linear
map $f\longmapsto\calo_f$ from functions $f\in C^\infty(\calm)$ to a
collection of symbols $\calo_f$ which close under some
$\bbc$-linear operation transporting the pointwise multiplication of
functions in a suitable sense; we leave the precise specification of these
operators intentionally vague for the moment, as their definition will depend on both the details of the magnetic Poisson
structure and on the specific quantization scheme that we adopt. The
minimal requirement is that they should reproduce the classical brackets
\eqref{eq:fgrho} at leading order in a deformation parameter $\hbar$,
which in physical scenarios we would like to identify with Planck's
constant of quantum mechanics. In other words, we demand that the
corresponding commutator brackets of the symbols satisfy
\begin{align}
[\calo_f,\calo_g]=\ii\hbar\, \calo_{\{f,g\}_\rho} + O(\hbar^2) \ ,
\label{eq:semiclassical}\end{align}
which just mimicks the correspondence principle of quantum
mechanics. Note that the quantization map is only required to be a
homomorphism of the bracket algebras to leading order in $\hbar$;
even in the simplest physical examples of canonical quantization, the
homomorphism property is violated in general at order $\hbar^2$. An
exception is the restriction of the map to linear (or quadratic)
polynomials in the coordinate functions of $C^\infty(\calm)$, which
preserves the fundamental brackets \eqref{eq:coordrho}:
\begin{align}
{}[\calo_{x^i},\calo_{x^j}]&=0 \ , \nonumber \\[4pt]
{}[\calo_{x^i},\calo_{p_j}]&=\ii\hbar\, \delta^i{}_j\,1_\calo \ ,
                     \label{eq:calocoord}      \\[4pt]
{}[\calo_{p_i},\calo_{p_j}]&=-\ii\hbar\, \rho_{ij}(\calo_x) \ , \nonumber 
\end{align}
where $1_\calo=\calo_1$ is the image of the constant unit function $1$
on $\calm$ under the quantization map.
This generally defines a nonassociative
algebra whose non-vanishing Jacobiators are given by
\begin{align}
[\calo_{p_i},\calo_{p_j},\calo_{p_k}] = -\hbar^2\, H_{ijk}(\calo_x)\, 1_\calo \ .
\label{eq:caloJac}\end{align}
The symbols $\rho_{ij}(\calo_x)$ and $H_{ijk}(\calo_x)$ 
depend in general on a choice of ordering, which is ambiguous; even in the simplest
examples of standard canonical quantization, different ordering prescriptions
lead to different quantization maps.

Whatever the quantization scheme, a fundamental ingredient in the
quantization of magnetic Poisson structures consists of the
\emph{magnetic translation operators}
\begin{align}
\calp_v=\exp\Big(\frac\ii\hbar\, \calo_{p\cdot v}\Big) \ ,
\label{eq:magtransloperators}\end{align}
defined symbolically here as global symbols via formal power series
expansions for fixed configuration space vectors $v=(v^i)\in
M=\bbr^d$. Their significance is that their conjugation action on the
symbols $\calo_f$ implements an action of the translation group
$T=\bbr^d$; indeed, a formal implementation of the first two commutation
relations in \eqref{eq:calocoord} yields
\begin{align}
\calp_v^{-1}\, \calo_{x^i}\,\calp_v=\calo_{x^i+v^i}
\label{eq:magtranslations}\end{align}
on the basic symbols corresponding to the configuration space
coordinates. This is expected on physical grounds from the
example of a charged particle in a magnetic field considered in
Section~\ref{sec:monopole}: A non-zero background magnetic field
breaks the exact translational symmetry of the classical dynamical system, but a remnant
of this symmetry remains after quantization in the form of a
``projective'' representation of the translation group $T=\bbr^d$, in
a suitable sense that we will make precise in the following. In
ordinary quantum mechanics, a group which is only represented
projectively is still a quantum symmetry.

\begin{figure}
\begin{center}
\includegraphics[height=14ex]{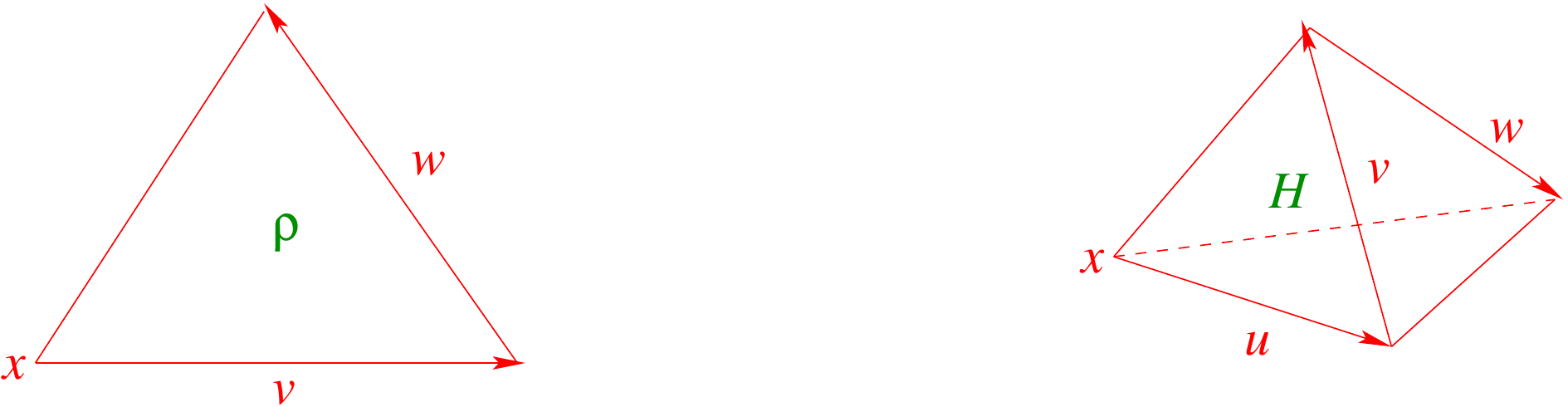}
\end{center}
\caption{The two-simplex $\triangle^2(x;v,w)$ on the left, over which
  the magnetic field
  $\rho$ is integrated, and the three-simplex $\triangle^3(x;u,v,w)$
  on the right, over which the magnetic charge $H$ is integrated.}
\label{fig:23simplex}\end{figure}

In fact,
proceeding again formally, by iterating \eqref{eq:magtranslations}
infinitesimally 
using the commutation and association relations \eqref{eq:calocoord} and
\eqref{eq:caloJac}, and then integrating,
we arrive at the corresponding global relations
\begin{equation}
\begin{aligned}
\calp_w\,\calp_v&=\e^{\ii{\Phi_2(x;v,w)}}\, \calp_{v+w} \ ,  \\[4pt]
\calp_w\,(\calp_v\,\calp_u) &= \e^{\ii{\Phi_3(x;u,v,w)}}\,
                              (\calp_w\,\calp_v)\,\calp_u \ .
\label{eq:magtranslrels}
\end{aligned}
\end{equation}
Here $\Phi_2(x;v,w)$ is given by the integral of the two-form $\rho$ through
the oriented two-simplex $\triangle^2(x;v,w)$ based at $x$ and spanned by the
translation vectors $v$ and $w$ in $M=\bbr^d$, while $\Phi_3(x;u,v,w)$
is given by the integral of the three-form $H=\dd \rho$ through the
oriented three-simplex $\triangle^3(x;u,v,w)$ based at $x$ and spanned by $u$,
$v$ and $w$ (see Figure~\ref{fig:23simplex}). In the $d=3$ example from
Section~\ref{sec:monopole}, the quantity $\Phi_2$ is interpreted as
the `magnetic flux' through the triangle $\triangle^2$ and $\Phi_3$ as the
`magnetic charge' enclosed by the tetrahedron $\triangle^3$. Since $ H=\dd\rho$, we
expect that the phase factor $\Phi_3$ defines a (trivial) three-cocycle in a certain
group cohomology of the
translation group $T$. These naive considerations go back
to~\cite{Jackiw:1984rd} (see also~\cite{Jackiw:2002wf}) in the context
of nonassociativity of symmetry operations in quantum field theory,
which lead to anomalies. 
Although this suggestion turns out to be
heuristically correct in the field of a magnetic monopole,
it has only been recently established in full generality at a rigorous level through
considerations of higher quantization. In particular, in the following
we shall address the following open questions which were not addressed
in the original treatment of~\cite{Jackiw:1984rd}:
\smallskip
\begin{enumerate}[i)]
\item How does one properly define the magnetic translation operators
  \eqref{eq:magtransloperators}?
\item What is the precise definition of the representation of the
  translation group $T=\bbr^d$ given by \eqref{eq:magtranslrels}?
\end{enumerate}

\subsection{Quantization I: \ $H=0$\label{sec:H=0}}

The questions posed above have well-known geometric answers in the
case that $\dd\rho=0$, which we
shall now review. In this case the two-form $\rho=\dd A$ can be written in
terms of a global one-form $A\in\Omega^1(M)$ on configuration space,
called a `vector potential', and we may identify $\rho=F_{\nabla^L}$
as the curvature of a connection $\nabla^L$ on a (trivial) complex line bundle $L$ over $
M=\bbr^d$. In this simple case the geometrical interpretation in terms
of line bundles is somewhat superfluous, but it has
the advantage that the treatment will generalize below to the more
complicated situations we are ultimately interested in. In this case the magnetic Poisson brackets
\eqref{eq:coordrho} generate an associative algebra and they can be
represented by the (unbounded) operators
\begin{equation}
\begin{aligned}
\calo_x&=x \ , \\[4pt]
\calo_p&=-\ii\hbar\,\nabla^L=-\ii\hbar\, \dd + A \ ,
\end{aligned}
\end{equation}
acting on the quantum Hilbert space $\hil={\rm L}^2(M,L)$ of
square-integrable sections of $L$ (equivalently square-integrable
functions on $M$ in this case). This is of course just the usual
prescription in standard geometric quantization of the symplectic manifold
$(\calm,\sigma_\rho)$, with polarization given by the integrable
distribution $TM$ which foliates the phase space $\calm$ by leaves
which are the Lagrangian
submanifolds $M\hookrightarrow\calm$, the image of the zero section of
the cotangent bundle $\calm=T^*M$.

\begin{figure}
\begin{center}
\includegraphics[height=12ex]{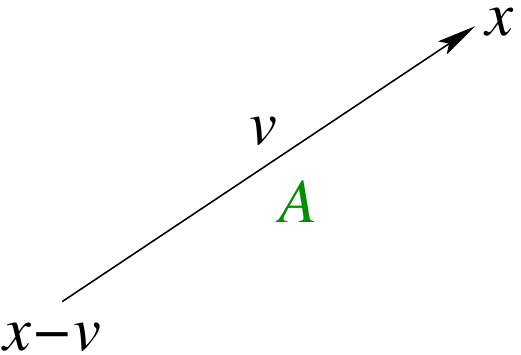}
\end{center}
\caption{The one-simplex $\triangle^1(x;v)$ along which
  the vector potential
  $A$ is integrated.}
\label{fig:1simplex}\end{figure}

Magnetic translations in this formulation have a natural geometric
definition as parallel transport in the line bundle $L$: For any
section $\psi\in\hil$, the magnetic translation operators
\eqref{eq:magtransloperators} can be defined by
\begin{align}
(\calp_v\psi)(x)=\exp\Big(-\frac\ii\hbar\, \int_{\triangle^1(x;v)}\,
  A\Big) \, \psi(x-v) \ ,
\label{eq:magtransldrho0}\end{align}
where $x\in M$, $v\in T$, and $\triangle^1(x;v)$ is the oriented
one-simplex based at $x$ along the vector $v$, i.e. the straight line
from $x-v$ to $x$ (see
Figure~\ref{fig:1simplex}). Explicit computation using $\dd A=\rho$
and Stokes' Theorem shows that they
define a (trivial) \emph{weak projective representation} of the translation group
$T=\bbr^d$ on the Hilbert space $\hil$ with
\begin{align}
(\calp_w\,\calp_v\psi)(x)=\omega_{v,w}(x) \ (\calp_{v+w}\psi)(x) \ ,
\end{align}
where
\begin{align}
\omega_{v,w}(x) = \exp\Big(-\frac\ii\hbar\,\int_{\triangle^2(x;w,v)}\,
  \rho\Big) \ . 
\label{eq:2cocycleweak}\end{align}
The adjective `weak' refers to the fact that \eqref{eq:2cocycleweak}
satisfies a twisted form of the usual cocycle condition given by
\begin{align}
\omega_{v,w}(x-u)\, \omega_{u+v,w}^{-1}(x)\, \omega_{u,v+w}(x)\,
  \omega_{v,w}^{-1}(x)=1 \ ,
\end{align}
and so defines a two-cocycle on the translation group $T$ with values in
$C^\infty(M,{\rm U}(1))$, the ${\rm U}(1)$-valued functions on $M$. In the special
case where the magnetic field $\rho$ is constant, i.e. the
component functions $\rho_{ij}$ are constant, the two-cocycle
\eqref{eq:2cocycleweak} simplifies to the constant phase
\begin{align}
\omega_{v,w}=\e^{-\frac{\ii}{2\hbar}\, \rho(v,w)} \ ,
\end{align}
and so defines a two-cocycle in the group cohomology $H^2(T,{\rm U}(1))$; in
this instance, the magnetic translation operators generate an ordinary
projective representation of $T$ on $\hil$.

In this class of magnetic Poisson structures it is also possible to
give the quantization map
$f\longmapsto\calo_f$ explicitly in terms of the 
\emph{magnetic Weyl transform} which sends a phase space function
$f\in C^\infty(\calm)$ to an operator $\calo_f\in{\rm End}(\hil)$. For
this, one introduces the magnetic Weyl system which is the family of
bounded operators\footnote{Here we are glossing over many technical
  functional analytic details: The domain of the magnetic Weyl
  transform is smaller than the smooth functions $C^\infty(\calm)$ and
  the Weyl operators $W(x,p)$ only act on a Schwartz subspace of
  $\hil$ in a suitable way; see e.g.~\cite{Bunk:2018qvk} for the
  precise treatment. These details are not important for the
  discussion that follows.} 
\begin{align}
W(x,p):\hil\longrightarrow\hil
\end{align}
parameterized by points $X=(x,p)\in\calm$ and defined using the magnetic translation operators by
\begin{align}
\big(W(x,p)\psi\big)(y)=\e^{\frac{\ii\hbar}2\,p\cdot x}\, \e^{-\ii
  p\cdot y} \, (\calp_x\psi)(y) \ .
\end{align}
The quantization map is then defined by the bounded operator
\begin{align}
\calo_f=\int_\calm\, \bigg( \int_\calm\, \e^{\ii\sigma_0(X,Y)}\, f(Y)
  \ \frac{\dd Y}{(2\pi)^d}\bigg)\, W(X) \ \frac{\dd X}{(2\pi)^d} \ ,
\end{align}
where $\dd X$ is the Lebesgue measure on $\calm$. This defines an
invertible linear transformation and the pre-image of the operator
product $\calo_f\,\calo_g$ defines the
\emph{magnetic Moyal--Weyl star product} $f\star_\rho g$ on
$C^\infty(\calm)$ through $\calo_{f\star_\rho
  g}:=\calo_f \,\calo_g$; explicitly, it can be written in terms of
the two-cocycle \eqref{eq:2cocycleweak} as an oscillatory integral
\begin{align}
&(f\star_\rho g)(X)=\notag \\ & \qquad =\frac1{(\pi\,\hbar)^d}\, \int_\calm \, \int_\calm
  \, \e^{-\frac{2\ii}\hbar\,\sigma_0(Y,Z)} \,
  \omega_{x+y-z,x-y+z}(x-y-z)\notag \\ & \hspace{4cm} \times \ f(X-Y)\, g(X-Z) \ \dd Y \ \dd Z \ .
\end{align}
This is a convergent star product (on a suitable algebra of Schwartz functions) and so defines a strict deformation
quantization of the magnetic Poisson structure on $\calm$. In the case
that $\rho$ is constant, it can be simplified to the usual Moyal--Weyl
form of a twisted convolution product
\begin{align}
&(f\star_\rho g)(X)= \label{eq:Moyalrho} \notag\\ &=\frac1{(\pi\,\hbar)^d}\, \int_\calm \, \int_\calm
  \, \e^{-\frac{2\ii}\hbar\,\sigma_\rho(Y,Z)} \, f(X-Y)\, g(X-Z) \ \dd
  Y \ \dd Z \ . 
\end{align}
This demonstrates another importance of the magnetic translation
operators: They provide a bridge between
\emph{geometric quantization} and \emph{deformation
  quantization}. This is the starting point for a reformulation of
canonical quantum mechanics as phase space quantum mechanics, see
e.g.~\cite{Szabo:2017yxd} and~\cite{Szabo:2018hhh} for reviews in the contexts of
magnetic monopole physics and non-geometric string theory, respectively.

\subsection{Quantization II: \ $H\neq0$\label{sec:Hnot0}}

Let us now turn to the quantization problem in our main case of
interest, when $H=\dd\rho\neq0$. In this case, the technical and
conceptual problem is that the operator/state
formulation of geometric quantization discussed above cannot handle
nonassociative magnetic Poisson algebras: Operators which act on a
separable Hilbert space always associate. 

The exception is the case of
a singular point distribution of magnetic charge. Recalling that the
magnetic field \eqref{eq:Diracmonopole} of a Dirac monopole is defined on
$M^\circ:=\bbr^3\setminus\{\vec 0\}$, we can consider the quantization
problem on the topologically non-trivial domain $M^\circ$ outside the
support of the magnetic charge distribution; physically, this is
tantamount to saying that the electric charge never reaches, and the
wavefunctions vanish at, the location of the monopole. In this case the magnetic
Poisson algebra is associative on $ M^\circ$, and the magnetic field
$\rho=\frac e\hbar \,\dd A_{\rm D}$ admits locally defined vector potentials
\eqref{eq:Diracpotential} which can be glued together by a gauge
transformation between the two 
Dirac string singularities along the unit vectors $\vec n$ and $-\vec
n$, corresponding
to the northern and southern hemispheres of the homotopy equivalent $S^2\simeq
M^\circ$. The two-form $\rho$ can then be
identified as the curvature of a connection $\nabla^L$ on a non-trivial line bundle
$L\longrightarrow M^\circ$ of first Chern class
\begin{align}
n = \frac{2\,e\,g}\hbar 
\label{eq:degree}\end{align}
if and only if $n\in\bbz$. This is the famous \emph{Dirac charge
  quantization} condition, and its present geometric derivation goes
back to~\cite{Wu:1976ge}. Then the quantum Hilbert space of geometric
quantization is $\hil={\rm L}^2(M^\circ,L)$. The magnetic Weyl
transform in this instance on
$\calm^\circ=T^*M^\circ$ can be constructed as before using magnetic
translation operators defined through parallel transport in the line
bundle $L$, which associate due to the quantization condition
\eqref{eq:degree}, and as previously it induces an associative phase space
star product~\cite{Soloviev:2017nwk}. Hence in this case too we find
an explicit form for the quantization map $f\longmapsto\calo_f$,
and hence a
natural correspondence between geometric quantization and deformation
quantization approaches.

On the other hand, for generic smooth distributions $H\in\Omega^3(M)$,
standard geometric quantization breaks down. Even if the support of
$H$ is a compact submanifold of $M$, one cannot consider the
quantization problem on the complement $M^\circ$ in the way we did above: The
quantization condition on the first Chern class \eqref{eq:degree} would be
violated by continuous deformations of $H$ in its support, and
hence no quantum line bundle for geometric quantization exists in this
case. In particular, for a uniform distribution of magnetic charge throughout
$M$, removing the support would leave $M^\circ=\emptyset$. Hence we
must resort to alternative approaches to the quantization of generic
magnetic Poisson structures. In the remainder of this contribution we
discuss three such perspectives on this quantization problem, and how
they address the questions raised in
Sections~\ref{sec:monopole}--\ref{sec:Magtransl} above.

\section{Perspective I: deformation quantization\label{sec:PI}}

The most straightforward way to quantize a general magnetic Poisson
structure is via Kontsevich's formalism for deformation
quantization of Poisson manifolds~\cite{Kontsevich:1997vb}, which can
be carried out without any reference to a Hilbert space formulation. At the
heart of Kontsevich's formalism is the Formality Theorem, which
provides an $L_\infty$-quasi-isomorphism from the $L_\infty$-algebra of
multivector fields, equipped with the Schouten bracket, to the
$L_\infty$-algebra of multidifferential operators, equipped with the
Gerstenhaber bracket. This
formalism can also be extended to quantize twisted Poisson structures
using algebroid
stacks~\cite{Kontsevich:2006ds,Severa:0205294,Aschieri:2002fq}. 

\begin{figure}
\begin{center}
\includegraphics[height=11ex]{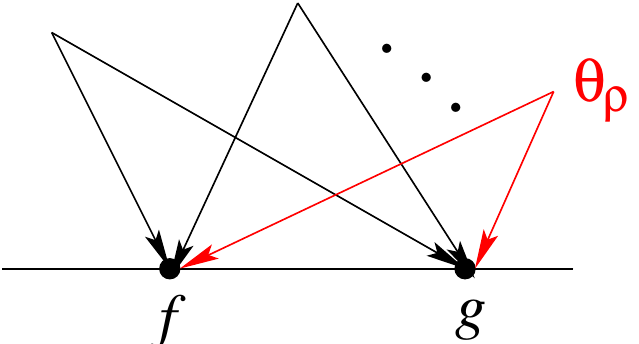}
\end{center}
\caption{Graphical representation of the contributions to ${\rm
    b}_n(f,g)$. There are $n$ insertions of the bivector $\theta_\rho$
in each diagram.}
\label{fig:bidiff}\end{figure}

In the setting of magnetic Poisson structures, which
was first worked out explicitly
in~\cite{Mylonas:2012pg}, 
Kontsevich's formality maps provide a noncommutative and
nonassociative star product on the algebra of formal power series
$C^\infty(\calm)[[\hbar]]$ for
any twisting three-form $H=\dd\rho\in\Omega^3(M)$. It is given by
\begin{align}
f\star_H g = f\, g + \frac{\ii\hbar}2\, \{f,g\}_\rho +
\sum_{n\geqslant2}\, \frac{(\ii\hbar)^n}{n!} \ {\rm b}_n(f,g) \ ,
\label{eq:starH}\end{align}
where ${\rm b}_n=U_n(\theta_\rho,\dots,\theta_\rho)$ are
bidifferential operators determined by the bivector
$\theta_\rho$. Here $U_n$ denote the formality maps which generally send a
collection of $n$ multivector fields to multidifferential operators,
and the functions ${\rm b}_n(f,g)$ can be computed by a combinatorial algorithm which sums over
graphs with prescribed weight given by suitable integrals over the
upper hyperbolic half-plane $\mathbb{H}$, that are represented
diagrammatically by applying the legs of the bivector insertions
$\theta_\rho$ to the functions $f$ and $g$ sitting on the boundary
$\bbr$ of $\mathbb{H}$ (see Figure~\ref{fig:bidiff}). 

\begin{figure}
\begin{center}
\includegraphics[height=11ex]{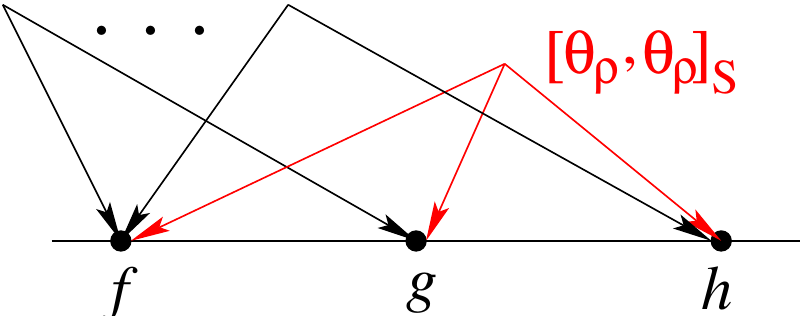}
\end{center}
\caption{Graphical representation of the contributions to ${\rm
    t}_n(f,g,h)$. There is a single insertion of the trivector $[\theta_\rho,\theta_\rho]_{\rm
  S}$ and $n-1$ insertions of the bivector $\theta_\rho$ in each
diagram.}
\label{fig:tridiff}\end{figure}

In addition to
providing a deformation quantization of the classical magnetic Poisson
brackets \eqref{eq:fgrho} in the sense of \eqref{eq:semiclassical}, the Kontsevich formality
maps further give a deformation quantization of the three-bracket defined by the
classical Jacobiator \eqref{eq:fghrho}. This quantum three-bracket is
given by
\begin{align}
[f,g,h]_{\star_H} = -\hbar^2\, \{f,g,h\}_\rho + \sum_{n\geqslant3}\,
\frac{(\ii\hbar)^n}{n!} \ {\rm
  t}_n(f,g,h) \ ,
\end{align}
where ${\rm t}_n=U_{n+1}([\theta_\rho,\theta_\rho]_{\rm
  S},\theta_\rho,\dots,\theta_\rho)$ are 
tridifferential operators, and the contributions ${\rm t}_n(f,g,h)$
can be computed by an analogous combinatorial algorithm in terms of
the trivector $[\theta_\rho,\theta_\rho]_{\rm
  S}$ which is
depicted in Figure~\ref{fig:tridiff}.

In the general case, the star product \eqref{eq:starH} is an
asymptotic series in $\hbar$ and defines a formal deformation quantization of the
magnetic Poisson structure. It is difficult to compute in generality
beyond the first few non-trivial orders in $\hbar$. However, for $H$
constant, and with the choice of magnetic field $\rho_{ij}(x) \=
\frac13\, H_{ijk}\, x^k$, the expansion simplifies enormously: Most
diagrams vanish and all non-vanishing contributions factorize as
powers of a single graph~\cite{Mylonas:2012pg}. The resulting series can be formally summed
to all orders in $\hbar$
and rewritten using Fourier transformations as a twisted convolution product~\cite{Mylonas:2013jha}
\begin{align}
&(f\star_H g)(X)= \label{eq:starHconst}\notag \\ &= \frac1{(\pi\,\hbar)^d}\, \int_\calm \, \int_\calm \,
  \e^{-\frac{2\ii}\hbar\,\sigma_\rho(Y,Z)} \, f(X-Y)\, g(X-Z) \ \dd Y
  \ \dd Z \ . 
\end{align}
Note that this formula is formally identical to the associative Moyal--Weyl type
star product \eqref{eq:Moyalrho} which was written in the case of
constant $\rho$. Here $\rho$ is not constant and the formula defines a
convergent nonassociative star product (on a suitable algebra of
Schwartz functions), giving a strict deformation quantization of the
magnetic Poisson structure in this instance.

In this framework, the quantization map is
simply $\calo_f=f$ with the multiplication on
$C^\infty(\calm)[[\hbar]]$ given by the star product
\eqref{eq:starH}. The nonassociative magnetic translation operators
\eqref{eq:magtransloperators} are thus given by the functions
\begin{align}
\calp_v:=\e^{\frac\ii\hbar\,p\cdot v} \ .
\end{align}
For $H$ constant, an explicit computation using the exact formula
\eqref{eq:starHconst} shows that the higher projective representation
\eqref{eq:magtranslrels} of the translation group $T=\bbr^d$ is determined by a three-cocycle
$\omega_{u,v,w}$ in the group cohomology $H^3(T,{\rm U}(1))$ with
\begin{equation}
\begin{aligned}
\calp_v\star_{ H}\calp_w &= \Pi_{v,w}(x) \ \calp_{v+w} \ , \\[4pt]
\big(\calp_u\star_{ H}\calp_v\big) \star_{ H} \calp_w &=
\omega_{u,v,w} \ \calp_u \star_{  H} \big(
\calp_v\star_{ H}\calp_w \big) \ , 
\label{eq:defquantrels}\end{aligned}\end{equation}
where 
\begin{align}
\Pi_{v,w}(x)=\e^{-\frac{\ii}{6\hbar}\, H(x,v,w)}
\end{align}
is a two-cochain with constant coboundary
\begin{align}
\omega_{u,v,w} = \e^{\frac{\ii}{6\hbar}\, H(u,v,w)} \ .
\end{align}
This approach thus answers the questions raised at the end of
Section~\ref{sec:Magtransl}. The perspective on nonassociativity in
terms of three-cocycles in the group cohomology $H^3(T,{\rm U}(1))$
was emphasised
by~\cite{Jackiw:1984rd,Mylonas:2012pg,Bakas:2013jwa}. They
can also be interpreted as three-cocycles of a suitable quasi-Hopf
algebra in a cochain twist approach to deformation
quantization~\cite{Mylonas:2013jha}.

It can be shown that the phase space formulation of nonassociative
quantum mechanics in
this setting, for constant magnetic charge $H$ or constant $R$-flux,
is physically sensible and gives novel testable quantitative
predictions~\cite{Mylonas:2013jha} (see
e.g.~\cite{Szabo:2017yxd,Szabo:2018hhh} for reviews). This answers
the questions raised at the end of Sections~\ref{sec:monopole}
and~\ref{sec:Rflux} in this particular instance. On the other hand, the approach suffers from many
problems. For example, the quantization is formal in $\hbar$ for
  non-constant $H$, so that for generic distributions of magnetic
  charge the deformation parameter cannot be identified with the
  physical Planck constant; as usual, deformation quantization is not
  a genuine quantization from this perspective. Moreover, there are
  the usual technical and conceptual issues associated with phase
  space quantum mechanics (see
  e.g.~\cite{Szabo:2017yxd,Szabo:2018hhh}). Finally, the loss of the
  Leibniz rule for the commutator constructed using the nonassociative
  star product casts doubt on the existence of suitable integrals of
  motion which enable at least partial integrability of the dynamical
  system. In light of these
  drawbacks, we would like to seek a framework which avoids the
  deficiencies of deformation quantization and takes us closer to a
  framework akin to an operator/state correspondence in geometric
  quantization. This is the goal of our next two perspectives.

\section{Perspective II: symplectic realization\label{sec:PII}}

One way to approach the problem of a Hilbert space formulation for the
quantization of magnetic Poisson structures is to generalize the
well-known technique of symplectic realization in Poisson geometry,
which allows one to cast the quantization of generic Poisson manifolds
into the framework of standard geometric quantization of symplectic manifolds. A 
\emph{symplectic realization} of a Poisson structure $\theta$ on a
manifold $M$ is a symplectic manifold $(S,\Omega)$ together with a
surjective submersion $S\longrightarrow M$ which is a Poisson map. The
original local construction goes back
to~\cite{weinstein1983local}, while a global formulation is
given in~\cite{karasev1987analogues,coste1987groupoides} using the corresponding
symplectic groupoid which integrates the Poisson manifold $(M,\theta)$. A
global generalization integrating twisted Poisson structures in terms
of almost symplectic manifolds is given in~\cite{Cattaneo:2003fs}. 

A \emph{local symplectic realization} of the magnetic Poisson algebra
is constructed in~\cite{Kupriyanov:2018xji}. Concretely, it 
``doubles'' $\calm$ to an extended phase space $\mathcal{S}$ with local coordinates
$(x^i,\tilde x^i,p_i,\tilde p_i)$ using local Darboux coordinates
$(x^i,\pi_i)$ and $(\tilde x^i,\tilde\pi_i)$ with the generalized Bopp
shifts $p_i=\pi_i-\frac12\, \rho_{ij}(x)\, \tilde x^j$ and $\tilde
p_i=\tilde\pi_i$. Then the symplectic algebra of coordinate functions
\begin{align}
\{x^i,p_j\}&=\{\tilde x^i,p_j\}=\{x^i, \tilde p_j\}=\delta^i{}_j \ ,
             \nonumber \\[4pt]
\{p_i,p_j\}&=\rho_{ij}(x)+\frac12\,\tilde
             x{}^k\,\Big(\frac{\partial \rho_{ij}(x)}{\partial x^k}-H_{ijk}(x)\Big) \ ,
             \\[4pt]
\{p_i,\tilde p_j\}&=\{\tilde p_i,p_j\}=\frac12\,\rho_{ij}(x)\notag
\label{eq:symplreal}\end{align}
is by construction an associative algebra. The pullback of the
corresponding symplectic form $\Omega$ on $\cals$ by the zero
section of the projection $\cals\longrightarrow\calm$ coincides with
the almost symplectic form
$\sigma_\rho$. The integrability of this realization is not yet
understood, though it is natural to speculate that it may be related to
the integration of Lie bialgebroids to Poisson--Lie
groupoids discussed in~\cite{Mackenzie:9712012}. A local generalization to arbitrary quasi-Poisson
structures is discussed in~\cite{Kupriyanov:2018yaj}.

This local symplectic realization is
intimately related to the approach via deformation
quantization discussed in Section~\ref{sec:PI}. The quantization of the
algebra \eqref{eq:symplreal} on $C^\infty(\calm)$ via the Schr\"odinger polarization
\begin{subequations}
\begin{align}
\widehat{\tilde p}_i&=\ii\hbar\, \frac\partial{\partial x^i} \ ,
 \\[4pt]
\widehat{\tilde x}{}^i&=-\ii\hbar\, \frac\partial{\partial p_i}
\end{align}
\end{subequations}
coincides with the associative composition algebra of differential operators $\big({\rm
  Diff}(\calm)[[\hbar]],\circ_H\big)$ on the original
phase space $\calm$ which governs observables in nonassociative
quantum mechanics~\cite{Mylonas:2013jha}. Here for phase space functions
$f,g\in C^\infty(\calm)$, the composition product $\circ_H$ is defined through
\begin{align}
(f\circ_H g)\star_H \varphi:=f\star_H(g\star_H\varphi)
\end{align}
for arbitrary test functions $\varphi\in
C^\infty(\calm)$, and in general $f\circ_H g $ is not a function but a formal
  power series in $\hbar$ of differential operators~\cite{Kupriyanov:2018xji}.

One can now ask how to reduce the extended dynamical system (at both classical
  and quantum levels) in a consistent way so as to recover the
  original magnetic Poisson manifold. This can be analysed by
  introducing the ${\rm O}(d,d){\times}{\rm O}(d,d)$-invariant Hamiltonian given by
\begin{align}
{\rm Ham}=p_I\,\eta^{IJ}\,p_J \ ,
\label{eq:HamLorentz}\end{align}
where $(p_I)=(p_i,\tilde p_i)$ and
\begin{align}
\eta=\begin{pmatrix} 0 & 1_d \\ 1_d & 0  \end{pmatrix}
\end{align}
is the constant ${\rm O}(d,d)$-invariant metric. With this choice, Hamilton's equations of motion with the non-degenerate
                                              Poisson brackets
                                              \eqref{eq:symplreal}
                                              reproduces, in the $d=3$ 
                                              example of
                                              Section~\ref{sec:monopole},
                                              the correct Lorentz
                                              force law for the
                                              physical coordinates
                                              $(x,p)$. 
The formalism is thus reminiscent of an approach based on double field
                                              theory.
In contrast, however, it is not possible to impose here a section
constraint that eliminates the unwanted degrees of
freedom resulting from the doubling. In~\cite{Kupriyanov:2018xji} it is shown that a consistent
Hamiltonian reduction of the symplectic realization can eliminate the
auxiliary coordinates $(\tilde x,\tilde p)$ if and only if $H=0$:
There is \emph{no} polarization of the extended symplectic algebra
which is consistent with both the Lorentz force and the original
nonassociative magnetic Poisson algebra. 

The problem with the approach based on symplectic realization thus remains in the
physical meaning of the spurious degrees of freedom. At first glance
an interpretation may be given by looking at the dynamics in $d=3$
dimensions: For constant magnetic charge $H$, the Lorentz force is
equivalent to the equations of motion of an electric charge in the
background of a Dirac monopole field \eqref{eq:Diracmonopole} with
additional frictional forces~\cite{Bakas:2013jwa}. A dissipative
dynamical system requires the introduction of extra degrees of freedom
representing a reservoir if one wishes to conserve the total
energy. However, in the present case the Hamiltonian
\eqref{eq:HamLorentz} is automatically an integral of motion and so
there is no need to introduce new variables in order to conserve the
energy. It is therefore not clear what the extra coordinates $(\tilde
x,\tilde p)$ mean in the symplectic realization of the magnetic
Poisson structure. Moreover, the extra variables hide the interesting
consequences of nonassociativity (by construction), such as the
three-cocycles characterizing the higher projective representation
defined by magnetic translation operators;
see~\cite{Kupriyanov:2018xji} for a discussion of this point. We therefore need
to appeal to some sort of Hilbert space formalism which can tackle the
nonassociativity of magnetic Poisson brackets head on. This is the
topic of our third and final perspective.

\section{Perspective III: higher geometric quantization\label{sec:PIII}}

Our considerations above motivate the desire to deal with
nonassociative algebras directly, which can be achieved through 
\emph{higher structures}: Whereas nonassociativity forbids the
definition of magnetic translations as operators on a separable
Hilbert space, as linear representations of groups are always
associative, nonassociativity can occur when representing elements of
a group by endofunctors of a symmetric monoidal category; the multiplication law may
then close only up to a natural isomorphism and the natural isomorphisms in
turn may only be represented up to higher projective phases. In the following
we shall refer to these ``representations'' as `weak projective 2-representations'.
In other words, we capture
nonassociativity by working in a more general category than the 
category of vector spaces normally used in physics. We pursue this line
of approach by categorifying the framework for geometric quantization
discussed in Section~\ref{sec:H=0}: We replace Hilbert spaces of
sections of line bundles for $\dd\rho=0$ with
2-Hilbert spaces of sections of a suitable geometric object which
encodes non-trivial magnetic charge $H=\dd \rho\neq0$. Specifically,
we provide a
natural geometric definition of nonassociative magnetic
translations by realizing them as parallel transport functors on a
bundle gerbe $\cali_\rho$ canonically associated with the magnetic
field two-form $\rho$ on $\bbr^d$.

\subsection{Bundle gerbes}

We begin by discussing what the suitable higher version of a line
bundle should be in order to pursue our quantization scheme. Let
$\pi:Y\longrightarrow M$ be a surjective submersion over a manifold
$M$; for example, we can take $Y$ to be the total
space of an open cover of $M$. Then the $p$-fold fibre products
  $Y^{[p]}  := Y\times_M \cdots \times_M Y$ form a simplicial
  space with face maps $\pi_i:Y^{[p]}\longrightarrow Y^{[p-1]}$
  defined by omitting the $i$-th entry of a $p$-tuple for
  $i=1,\dots,p$; for example, $Y^{[p]}$ can be the space of $p$-fold
  intersections of open sets of a cover of $M$. For $p=2$ this defines the pair groupoid
  $Y^{[2]} \rightrightarrows Y$ whose 
  source and target maps are $\pi_2$ and $\pi_1$, respectively, and
  whose orbit space is the base manifold
  $M$ itself. Then a \emph{bundle gerbe} $(L,Y)$ on $M$ is a groupoid
  central extension of $Y^{[2]} \rightrightarrows
  Y$~\cite{Murray:9407015}; it can be depicted by
\begin{align}
\xymatrix@C=10mm{
& L \ar[d]_{\bbc} & \\
& Y^{[2]} \ \ar@< 2pt>[r]^{\pi_2} \ar@< -2pt>[r]_{\pi_1} & \ Y \ar[d]^\pi \\
 & & M
}
\label{eq:groupoidcentralext}\end{align}
Here we assume that $L$ is a complex line bundle over $Y^{[2]}$ in
order to extend our previous considerations, but the definition also
holds at the level of principal ${\rm U}(1)$-bundles. The groupoid
multiplication $(y_1,y_2)\circ(y_2,y_3)=(y_1,y_3)$ on $Y^{[2]} \rightrightarrows Y$ additionally gives a
bundle gerbe multiplication which is a bundle isomorphism $\mu:\pi_3^*(L)\otimes \pi_1^*(L) \xrightarrow{ \ \simeq \ } 
\pi_2^*(L)$ over $Y^{[3]}$, and can be depicted by
\begin{align}
\xymatrix@C=10mm{
\pi_3^*(L)\otimes\pi_1^*(L)
\xrightarrow{ \ \mu \ } \pi_2^*(L) \ar[d]
& L \ar[d] & \\
Y^{[3]} \ \ar@< 3pt>[r] \ar@<0pt>[r] \ar@< -3pt>[r]_{\pi_i} & Y^{[2]} \ \ar@< 2pt>[r]^{\pi_2} \ar@< -2pt>[r]_{\pi_1} & \ Y \ar[d]^\pi \\
 & & M
}
\end{align}
The bundle gerbe multiplication is associative over $Y^{[4]}$.

A \emph{connection} on a bundle gerbe $(L,Y)$ is a connection
$\nabla^L$ on the
line bundle $L\longrightarrow Y^{[2]}$ together with a two-form $\rho\in\Omega^2(Y)$ satisfying 
  $\pi_2^*(\rho)-\pi_1^*(\rho)=F_{\nabla^L}$. By the Bianchi identity
  $\dd F_{\nabla^L}=0$, the three-form $\dd\rho$ on $Y$ descends to a
  closed three-form $H\in\Omega^3(M)$, with $\pi^*H=\dd\rho$, which is
  called the curvature of the bundle gerbe. Analogously to the first
  Chern class for line bundles, the curvature $H$ is a de~Rham
  representative of a higher characteristic class in $H^3(M,\bbz)$
  measuring obstructions to topological triviality of the bundle gerbe (in a
  suitable sense that we do not spell out here), called the
  Dixmier--Douady class. It is clear then that bundle gerbes with
  connection are the appropriate receptacle to describe a geometric
  approach to the quantization of generic magnetic Poisson structures.

\subsection{Sections of bundles gerbes}

To define sections of bundle gerbes, we recall that Hermitian vector
bundles (with connection) on $M$ are objects in a symmetric monoidal category $\text{HVbdl}(M)$
under tensor product of vector bundles; under the direct sum of vector
bundles, it moreover has the structure of a rig category, the
categorification of a ring without additive inverses. A section of a vector bundle
$V\longrightarrow M$ can be equivalently regarded as a morphism in
this category from the trivial vector bundle $I_0$ with connection to $V$, and in particular a
morphism from the trivial bundle $I_0$ to itself is the same thing as a function
on $M$; the $C^\infty(M)$-module structure of $\Gamma(M,V)$ translates
into an action of the latter space of morphisms on the former via
composition of morphisms. Under the embedding of $\bbc$ into the morphisms as the
constant functions on $M$, this makes $\text{HVbdl}(M)$ into a
$\bbc$-linear category. 

Analogously, bundle gerbes $\cG=(L,Y)$ (with
connection) on $M$ are objects in a
symmetric monoidal 2-category. The original construction is due
to~\cite{Waldorf:0702652}, and was subsequently extended
by~\cite{Bunk:2016rta} to show that the 2-category in question has the
further structure of a closed abelian symmetric monoidal category enriched in symmetric
monoidal categories. A \emph{section} of a bundle gerbe $\cG=(L,Y)$ is
is a (left) module over $\cG$, defined by a vector bundle $E$ over $Y$
and an action $L\otimes\pi_1^*E\longrightarrow\pi_2^*E$ which is an
isomorphism of bundles over $Y^{[2]}$ satisfying the
obvious associativity constraint on $Y^{[3]}$; this is the same thing
as a 1-morphism from the trivial bundle gerbe
$\cali_0$ with connection to $\cG$. Then the \emph{2-Hilbert space of sections
  $\Gamma(M,\cG)$} is defined to be the $\text{Hilb}$-module category
of morphisms $\cali_0\longrightarrow\cG$, where $\text{Hilb}$ is the symmetric monoidal
category of finite-dimensional complex Hilbert spaces under tensor product.\footnote{The
  restriction to finite-dimensional Hilbert spaces and vector bundles
  of finite rank restricts to bundle gerbes $\cG$ which have torsion
  Dixmier--Douady class. This restriction will be irrelevant for our
  considerations below, as we shall always work with trivial bundle
  gerbes over $M=\bbr^d$. See~\cite{Bunk:2016rta} for a more in-depth
  discussion of this issue.} The category $\Gamma(M,\cG)$ enjoys the
following properties:
\smallskip
\begin{enumerate}[i)]
\item It carries the structure of a rig module category over
  the rig category $\text{HVbdl}(M)$.
\item There is an inner product bifunctor $\langle \ , \ \rangle:\Gamma(M,\cG)^{\rm op}\times\Gamma(M,\cG)\longrightarrow \text{Hilb}$.
\end{enumerate}
\smallskip
See~\cite{Bunk:2016rta,Bunk:2016gus} for precise definitions and
further details. In this categorification the ground field $\bbc$ is
replace with the rig category $\text{Hilb}$ and the ring of functions
$C^\infty(M)$ is replaced with the rig category $\text{HVbdl}(M)$. 

The 2-Hilbert space $\Gamma(M,\cG)$ admits a particularly simple
description on $M=\bbr^d$, in which case we can restrict attention to
topologically trivial bundle gerbes $\cG\=\cali_\rho$ with connection
specified entirely by a globally defined two-form $\rho\in\Omega^2(M)$. The groupoid
central extension \eqref{eq:groupoidcentralext} can then be replaced
with the simpler extension
\begin{align}
\xymatrix@C=10mm{
M\times\bbc \ar[d] & \\
M \ \ar@< 2pt>[r] \ar@< -2pt>[r] & \ M \ar[d]^{\rm id} \\
 & M
}
\label{eq:simplerext}\end{align}
and up to equivalence the category $\Gamma(M,\cG)$ admits the following concrete
description:\footnote{This is not a generic description of the
  2-Hilbert space in the topologically trivial case $[H]=0$, but
  rather of a full subcategory defined by the trivializations of the
  line bundle $L\longrightarrow Y^{[2]}$ and surjective submersion
  $\pi:Y\longrightarrow M$ indicated in \eqref{eq:simplerext}, which
  is sufficient for the present purposes; see~\cite{Bunk:2018qvk} for
  a more detailed discussion of this point. For notational ease, we
  continue to use the same symbol $\Gamma(M,\cali_\rho)$ for this subcategory.} 
  \smallskip
\begin{enumerate}[i)]
\item Its objects are topologically trivial Hermitian vector bundles
  with connection on $M$, i.e. globally defined one-forms
  $\eta\in\Omega^1(M,{\rm u}(n))$ valued in Hermitian $n\times n$
  matrices for any $n\in\mathbb{N}_0$.
\item Its morphisms are parallel morphisms of vector bundles with
  connection on $M$, i.e. a morphism $f:\eta\longrightarrow\eta'$ is
  an $n\times n'$ matrix-valued function $f:M\longrightarrow {\rm
    Mat}(n\times n')$ satisfying $\ii\eta'\,f=\ii f\,\eta-\dd f$. This resembles a gauge transformation except that $f$ need not be
  invertible: In the construction of the 2-Hilbert space the
  non-invertible 1-morphisms are essential, and these correspond to the
  matrix-valued functions of size $n \times n'$ with $n\neq n'$.
\end{enumerate}

\subsection{Magnetic translation functors}

A precise definition of nonassociative magnetic translations was
constructed in~\cite{Bunk:2018qvk} on the 2-Hilbert space
$\Gamma(M,\cali_\rho)$ in terms of the parallel transport functor
$\calp_v:\Gamma(M,\cali_\rho)\longrightarrow \Gamma(M,\cali_\rho)$
which is defined on objects $\eta$ as an infinitesimal version of the
parallel transport operators \eqref{eq:magtransldrho0} and by the
usual action of the translation group $T=\bbr^d$ on morphisms $f$ via
pullback $v^*(f)$:
\begin{subequations}
\begin{align}
\calp_v(\eta)|_x(a)&=\eta|_{x-v}(a)+\frac1\hbar\,
                     \int_{\triangle^1(x;v)} \, \iota_a\rho \ ,
  \\[4pt]
\calp_v(f)(x)&=f(x-v) \ ,
\end{align}
\end{subequations}
where $v\in T$, $x\in M$ and $\iota_a$ denotes contraction with the
vector $a\in\bbr^d$. This definition can be understood by transgressing
the gerbe $\cali_\rho$ to a line bundle with connection over the loop
space $LM$ of $M$~\cite{Bunk:2016rta}, and defining parallel transport
over $LM$ in the usual way.
That is, $\calp_v$ translates a one-form $\eta$ by $v$ and adds the one-form obtained by  
integrating $\rho$ along the one-simplex $\triangle^1(x;v)$; the
extra term is necessary in the parallel condition in order to induce an additional
one-form coming from the integration of $H= \dd\rho$ over a
two-simplex, which is the incarnation of the transgression line bundle over the  
boundary of a two-simplex relating parallel transport along two  
paths with the same endpoints. This defines a weak module functor in the sense that
\begin{align}
\calp_v(\xi\otimes\eta)= v^*(\xi)\otimes\calp_v(\eta) \ ,
\end{align}
for $\xi\in\Omega^1(M,{\rm u}(k))$.

The magnetic translation functors are subjected to natural coherence
isomorphisms
\begin{subequations}
\begin{align}
\Pi_{v,w}:\calp_v\circ\calp_w\Longrightarrow \chi_{v,w}\otimes\calp_{v+w}
\label{eq:magcoh}\end{align}
where
\begin{align}
\chi_{v,w}|_x(a)=\frac1\hbar\,\int_{\triangle^2(x;w,v)} \, \iota_aH 
\end{align}
is a connection one-form of the trivial line bundle on $M$, which corresponds to the transgression line bundle over
the two-simplex $\triangle^2(x;w,v)$. The
components of $\Pi_{v,w}$ are defined by
\begin{align}
\Pi_{v,w|\eta}(x):= \exp \Big(-\frac\ii\hbar\,
  \int_{\triangle^2(x;w,v)}\, \rho \Big) \ .
\end{align}
\end{subequations}

`Nonassociativity' is a natural property of the
coherence isomorphisms when applied to the two possible bracketings of a triple
composition of parallel transport functors
$\calp_u\circ\calp_v\circ\calp_w$. By iterating \eqref{eq:magcoh}, this identifies natural
transformations through
\begin{subequations}
\begin{align}
\Pi_{u+v,w}\circ\Pi_{u,v}(x) = \omega_{u,v,w}(x) \  \Pi_{u,v+w}\circ\calp_u(\Pi_{v,w})(x)
\end{align}
where
\begin{align}
\omega_{u,v,w}:\chi_{u+v,w}\otimes\chi_{u,v}\longrightarrow
\chi_{u,v+w}\otimes u^*(\chi_{v,w})
\end{align}
is the morphism in
$\Gamma(M,\cali_\rho)$ defined by
\begin{align}
\omega_{u,v,w}(x):=\exp\Big(\frac\ii\hbar\,
  \int_{\triangle^3(x;w,v,u)}\, H\Big) \ .
\end{align}
\end{subequations}
It is these coherence identities which make precise the relations
\eqref{eq:magtranslrels} among magnetic translations, and we shall
discuss their precise representation theoretic meaning below. In the
case that the magnetic charge $H$ is constant, the pertinent
${\rm U}(1)$-valued functions on $M$ simplify to 
\begin{subequations}
\begin{align}
\Pi_{v,w|\eta}(x)&=\e^{-\frac\ii{6\hbar} \, H(x,v,w)} \ , \\[4pt]
\omega_{u,v,w}&=\e^{\frac\ii{6\hbar}\, H(u,v,w)} \ ,
\end{align}
\end{subequations}
which agrees with what we found in Section~\ref{sec:PI} in the
approach based on deformation quantization.

In Section~\ref{sec:PI} the relations \eqref{eq:defquantrels} showed
that $\Pi_{v,w}$ and $\omega_{u,v,w}$ (for $H$ constant) have natural interpretations in
the group cohomology $H^3(T,{\rm U}(1))$, and define a
``higher'' projective representation of the translation group $T$ in
this sense. Here we can make this notion of higher projective
representation more precise in the language of category theory. We do
not spell out the general definitions, which can be found
in~\cite{Bunk:2018qvk}, but simply use the properties discussed above
to characterize these notions. Analogously to the $H=0$ case, $\omega_{u,v,w}$ define a three-cocycle on
$\bbr^d$ with values in $C^\infty(M,{\rm U}(1))$. On the other hand, the
pairs $(\chi_{v,w},\omega_{u,v,w})$ define a \emph{higher weak
  two-cocycle} on $\bbr^d$ with values in the $\text{Hilb}$-algebra
category $\text{HVbdl}(M)$. These constructions were collected
together in~\cite{Bunk:2018qvk} to give the following central result.

\begin{theo} The pairs $(\calp_v,\Pi_{v,w})$ \ define a \emph{weak
    projective 2-representation} of the translation group $\bbr^d$ on
  the $\text{HVbdl}(M)$-module category $\Gamma(M,\cali_\rho)$, the
  2-Hilbert space of sections of the bundle gerbe $\cali_\rho$ on $M$.
\end{theo}

This approach thus completely answers the questions which arose at the
end of Section~\ref{sec:Magtransl} in full generality. However, at this stage it is not
clear how it addresses the issues raised in Sections~\ref{sec:monopole}
and~\ref{sec:Rflux}, and there are many open issues which remain. In
the framework of nonassociative quantum mechanics, it is not clear
what is the physical significance of the 2-Hilbert space
$\Gamma(M,\cali_\rho)$ in terms of states, observables, and so on;
see~\cite{Bunk:2018qvk} for some preliminary analysis along these
lines, where it is also shown how the fake curvature condition of
higher gauge theory
naturally emerges in terms of this interpretation when considering
covariant derivatives on these bundle gerbes. In this sense it would be interesting to develop a ``higher
magnetic Weyl transform'' to determine the quantization map as a
natural transformation from functions $f\in C^\infty(\calm)$, regarded
as objects in the functor category $[\calm\rightrightarrows\calm,
\bbc\rightrightarrows\bbc]$ between the discrete categories based on
the sets $\calm$ and $\bbc$, to objects $\calo_f$ in the functor
category $[\Gamma(M,\cG),\Gamma(M,\cG)]$. In particular, this would bridge the approach based on higher
geometric quantization with deformation quantization, extending what
was described in Section~\ref{sec:H=0} for the associative case
$H=0$ and in Section~\ref{sec:Hnot0} in the case of singular magnetic
charge distributions.

It is also an open issue as to whether or not the constructions
discussed here, which deal with topologically trivial bundle gerbes,
can be adapted to classes of non-trivial bundle gerbes, without the
use of a trivializing open cover. For example, it would be interesting to extend
the weak projective 2-representations to compact Lie groups using the
quantum field theory construction of three-cocycles
in~\cite{Mickelsson:2008xn}. In this case the definition of the
2-Hilbert space is more involved due to technical difficulties related
to the non-torsion gerbes~\cite{Bunk:2016rta}. However, at least the definition of a
higher weak two-cocycle should carry over.

\bibliography{allbibtex}

\bibliographystyle{prop2015}

\end{document}